\documentclass[fleqn,twoside]{article}%
\usepackage{amsmath}
\usepackage{amsfonts}
\usepackage{amssymb}

\usepackage{graphicx}%
\usepackage{caption}
\usepackage{cite}
\def\be{\begin{equation}}
\def\ee{\end{equation}}
\def\bi{\bibitem}

\begin{document}
\title{Reconstructing modified and alternative theories of gravity.}
\author{Dalia Saha$^{1,2}$, Manas Chakrabortty$^3$ and Abhik Kumar Sanyal$^4$}
\maketitle
\noindent
\begin{center}
\noindent
$^{1,4}$ Dept. of Physics, Jangipur College, Murshidabad, West Bengal, India - 742213,\\
$^{2}$ Dept. of Physics, University of Kalyani, Nadia, West Bengal, India - 741235.\\
$^{3}$ Dept. of Physics, University of Bankura, Bankura, West Bengal, India - 722155.\\
\end{center}
\footnotetext[1]{
\noindent Electronic address:\\
$^1,^2$daliasahamandal1983@gmail.com\\
$^3$ manas.chakrabortty001@gmail.com\\
$^4$sanyal\_ak@yahoo.com\\}
\noindent
\date{}
\maketitle
\begin{abstract}
A viable radiation dominated era in the early universe is best described by the standard (FLRW) model of cosmology. In this short review, we demonstrate reconstruction of the forms of F(R) in the modified theory of gravity and the metric compatible F(T) together with the symmetric F(Q) in alternative teleparallel theories of gravity, from different perspectives, primarily rendering emphasis on a viable FLRW radiation era. Inflation has also been studied for a particular choice of the scalar potential. The inflationary parameters are found to agree appreciably with the recently released observational data. \\
\end{abstract}

\section{Introduction:}

Luminosity versus red-shift data curve associated with SN1a standard candles requires a modification of FLRW universe at the late stage of cosmic evolution. On the other hand, the early vacuum dominated era requires to incorporate a stage of very rapid growth of the universe (Inflation) in order to alleviate the flatness, the horizon and the structure formation problems associated with the standard model of cosmology. However, the radiation dominated era and the early matter dominated era are best described by the standard model of cosmology. In this connection let us mention that FLRW universe or the FLRW model stands for the standard model of cosmology and must not be confused with the `FLRW metric'. In fact, based on the cosmological principle (homogeneity and isotropy), Friedman in 1922 and Lema$\mathrm{\hat{i}}$tre in 1927 independently showed that a universe filled with perfect barotropic fluid undergoes decelerated expansion ($a \propto \sqrt t$ in the radiation era and $a \propto t^{(2/3)}$ in the matter dominated era, where, $a(t)$ is the scale factor). Later, Robertson (1935, 1936) and Walker (1936) independently demonstrated that the metric used by Friedmann and Lema$\mathrm{\hat{i}}$tre provides the most general Riemannian geometry compatible with homogeneity and isotropy. After Hubble’s discovery in 1930 along with Eddington’s proof (1931) that the static universe with a positive cosmological constant ($\Lambda$) is unstable, Einstein withdrew $\Lambda$ and agreed that the universe is expanding. Since then it is dubbed as the standard model of cosmology or the FLRW model. However, for the reason stated above the metric which was commonly known as Robertson-Walker metric, currently called as FLRW metric. Note that the metric and the model are different. In the background of FLRW metric, if the energy momentum tensor is modified by considering any other type of fluid (dissipative, scalar field or even a cosmological constant) it is different from the FLRW model.\\

Since on one hand, the theoretical estimation of cosmological constant (the sum of zero point energy of all quantum fields existed in the early universe) is $10^{120}$ order of magnitude larger than the value required to solve the cosmic puzzle (late-stage of accelerated expansion) and on the other, no trace of dark energy is observed after performing precession experiments \cite{1}, therefore the bulk of recent research is oriented in search of alternatives. Consequently, a host of modified theories of gravity have been proposed in recent years as alternatives to the dark energy. However considering the fact that, apart from the violation of superposition principle, the modified $F(R)$ theories, being associated with higher derivative terms, sometimes run from Ostrogradsky's instability \cite{2}; alternative theories of gravity, commonly known as teleparallel theories are currently in the limelight. \\

Curvature is the trait of (attributed to the) `general theory of relativity' (GTR), where the torsion-less Levi-Civita connection is used. The minimal modified version of GTR is the so called $F(R)$ theory of gravity. Alternatively, it is possible to ascribe gravity to torsion, by using the curvature-less (${R^\alpha}_{\beta\mu\nu} = 0$) Weitzenb\"ock connection. Since $\nabla_\mu g_{\alpha\beta} = 0$, where $\nabla$ is the covariant derivative with respect to the affine connection instead of the Levi-Civita connection; therefore it is often referred to as metric (compatible) teleparallel gravity. Its extended version is the $F(\mathrm T)$ theory of gravity, where $\mathrm T$ is the quadratic torsion scalar constructed from the torsion tensor. Yet another class of gravity theory exists by the name symmetric teleparallel gravity. It is constructed from curvature-less (${R^\alpha}_{\beta\mu\nu} = 0$) and torsion-less (${\mathrm T^\lambda}_{\mu\nu} = 0$) general affine connection, which is symmetric in lower indices. It's generalized version is the $F(Q)$ gravity theory, where, $Q$ stands for the quadratic non-metricity scalar constructed from the non-metricity tensor ($Q_{\lambda\mu\nu} = \nabla_\lambda g_{\mu\nu}$). Clearly, in these so-called teleparallel theories of gravity, the affine connection is independent of the metric tensor and so these are essentially the metric-affine theories, where both the metric and the connection act as dynamical variables. To the first order, both the teleparallel theories end up with GTR apart from a total derivative term and are dubbed as `Teleparallel Equivalent of General Relativity' (TEGR) and `Symmetric Teleparallel Equivalent of General Relativity' (STEGR) respectively. However, their extended versions deviate considerably from GTR and its modified versions $F(R)$ or else, particularly giving second order differential field equations and thus avoiding Ostrogradsky's instability. However, $F(\mathrm T)$ gravity theory is endured with a serious pathology of non-invariance under local Lorentz transformations of the tetrad field. It may be stated that Lorentz symmetry is the only known fundamental physical symmetry. Lack of  Lorentz invariance requires preferred frames to parallelize the spacetime, implying the existence of extra degrees of freedom (dof) compared to GTR. The canonical formulation of the singular action of $F(\mathrm T)$ gravity theory requires constraint analysis. The extra dof appear due to the six first-class constraints associated with the theory \cite{2a, 2b}. Recent investigation suggests that the number of extra dof is one \cite{2c, 2d, 2e}. Next, is the problem endured with strong coupling issue. This appears due to the fact that some of the genuine physical degrees of freedom lose their kinetic term at the quadratic order and consequently, the standard perturbation theory breaks down on these backgrounds \cite{purt1, 3}. Under these circumstances, the extended version of symmetric teleparallel $F(Q)$ gravity theory, apparently being free from such problems \cite{4}, attracted lot of attentions in the recent years. Nonetheless, it is to be mentioned that, as long as the cosmic puzzle is not being resolved unambiguously, focus on different modified and alternative theories of gravity will persist.\\

Whatever might be the theory, whenever an arbitrary functional form like $F(R)$, $F(\mathrm{T})$ or $F(Q)$ is retained in the action, it becomes mandatory to expatiate such functional forms in view of some additional physically viable conditions. Two possibilities have largely been explored so far. One is: to invoke the presence of dynamical (Noether) symmetry, while the other: is to follow the reconstruction program. Both have come out with certain specific functional forms in all the cases. Additionally, some forms are also chosen by hand (on some physical ground). Most of these, claim to associate early inflation in the vacuum era together with late-time accelerated expansion in the current matter dominated era. The important point to be noticed is that the cosmic evolution in the radiation dominated era is not studied in general, as if the universe jumped from the early inflationary stage to the late stage of matter dominated era. On the contrary, a viable radiation dominated era which can explain the standard Big-Bang Nucleosynthesis (BBN), account for the observed structure formation and the formation of CMBR, is best described by the standard (FLRW) model of cosmology. This manuscript is devoted to reconstruct all the three different forms for the gravity theories with curvature [$F(R)$], torsion [$F(\mathrm T)$], as well as non-metricity [$F(Q)$] of different extended theories of gravity in view of a viable radiation dominated era.\\

In the following section 2, we briefly review the above mentioned three pilers of gravity theory. In section 3, we follow the reconstruction program as mentioned, to explore the forms of $F(R)$, and $F(\mathrm {T})$ gravity theories, primarily in view of a viable Friedmann-like radiation dominated era. Form of $F(\mathrm{T})$ is also known from Noether symmetry consideration, which is different from the form so obtained here. We therefore have studied the energy conditions and inflation taking into account different forms of $F(\mathrm {T})$ so obtained. Since, $F(Q)$ gravity in coincident gauge is no different from $F(\mathrm{T})$ in the present context, so all the results obtained for $F(\mathrm{T})$ theory holds as well in $F(Q)$ theory of gravity. Yet another largely developed theory of gravity, the so called Palatini formalism, has been discussed briefly at the end of section 3. Finally we conclude in section 4.

\section{Three pilers of Gravitation:}

The building block of gravitation is the general affine connection ${\Gamma^\alpha}_{\mu\nu}$, which is constructed from the Levi-Civita connection ($\{_\mu{^\alpha}_\nu\}$), the contorsion tensor (${K^\alpha}_{\mu\nu}$) and the disformation tensor (${L^\alpha}_{\mu\nu}$) as,

\be\label{2.1} {\Gamma^\alpha}_{\mu\nu} = \{_\mu{^\alpha}_\nu\}  + {K^\alpha}_{\mu\nu} + {L^\alpha}_{\mu\nu},\ee
where,
\be\label{2.2} \begin{split}&\{_\mu{^\alpha}_\nu\} = \frac{1}{2} g^{\alpha\lambda}\left(g_{\mu\lambda,\nu}+g_{\nu\lambda,\mu}-g_{\mu\nu,\lambda}\right),\\&
{K^\alpha}_{\mu\nu} = \frac{1}{2}g^{\alpha\lambda}\left(\mathrm {T}_{\lambda\mu\nu}-\mathrm{T}_{\mu\lambda\nu}-\mathrm{T}_{\nu\lambda\mu}\right),\\&{L^\alpha}_{\mu\nu} = \frac{1}{2}g^{\alpha\lambda}\left(Q_{\lambda\mu\nu}-Q_{\mu\lambda\nu}-Q_{\nu\lambda\mu}\right).\end{split}\ee
While, the Levi-Civita connection is formed from the derivatives of the metric, the contorsion tensor, being the difference between the connections with and without torsion, is again equivalent to the difference between the Weitzenb\"ock (${\Gamma^\lambda}_{a\mu} = e^{a\lambda}\partial_\mu e_{a\nu}$, where ${e^a}_\mu$ are the components of the orthonormal basis vectors of the tetrad field, commonly called the vielbein, or vierbein field, to be more specific. Here, Latin indices stands for Lorentz indices, while Greek indices stand for space time. Since both run from $0-3$, we shall omit using Latin indices in the following.)\footnote{It may be mentioned that the curvature tensor constructed out of Weitzenb\"ock connection vanishes identically, while a theory of gravity constructed in view of tetrad fields admits both the Weitzenb\"ock and the Riemannian geometry and therefore is more general.} and the Levi-Civita connections, and is formed from the torsion tensor as,
\be\label{C}{\mathrm {T}^\alpha}_{\mu\nu} = \partial_\mu {e^\alpha}_\nu - \partial_\nu {e^\alpha}_\mu = {\Gamma^\alpha}_{[\mu,\nu]} = {\Gamma^\alpha}_{\mu\nu} - {\Gamma^\alpha}_{\nu\mu},\ee
and finally, the disformation tensor is constructed from the non-metricity tensor $Q_{\alpha\mu\nu} = \nabla_\alpha g_{\mu\nu} \ne 0$ as,
\be\label{D} Q_{\alpha\mu\nu} = \nabla_\alpha g_{\mu\nu}= g_{\mu\nu,\alpha} -  {\Gamma^\lambda}_{\mu\alpha}g_{\lambda\nu} -  {\Gamma^\lambda}_{\nu\alpha}g_{\mu\lambda}.\ee
Einstein's GTR is built from the non-vanishing Riemann tensor formed from the Levi-Civita connection,

\be \label{R} {R^\alpha}_{\beta\mu\nu} = \partial_\mu\{_\nu{^\alpha}_\beta\} - \partial_\nu\{_\mu{^\alpha}_\beta\} + \{_\mu{^\alpha}_\lambda\}\{_\nu{^\lambda}_\beta\} - \{_\nu{^\alpha}_\lambda\}\{_\mu{^\lambda}_\beta\},\ee
the simple reason being, the Levi-Civita connections are the unique affine torsion-free connections on the tangent bundle of a manifold that preserves Riemannian metric ${g^{\mu\nu}}_{;\nu} = 0$. So once a metric is specified, the unique non-vanishing components of Riemann tensor are found from the Levi-Civita connections or Christoffel symbols to be specific\footnote{Under the choice of local coordinates and coordinate basis $x^\mu$ and $\partial_\mu$ respectively, the affine Levi-Civita connection ($\nabla$) is called the Christoffel symbol $\{_\mu{^\alpha}_\nu\}$ so that $\nabla_\mu \partial _\nu = \{_\mu{^\alpha}_\nu\}$.} \eqref{R}. The above relation \eqref{R} is contracted to find the Ricci tensor ($R_{\mu\nu}$) and under further contraction Ricci scalar ($R$) results in. While, the Einstein-Hilbert action for GTR is constructed from the Ricci scalar, the so called modified theory of gravity is constructed from $F(R)$ as,

\be\label{A1} A = \int\left[F(R)\sqrt{-g}~d^4 x  \right] + \mathcal{S}_m,\ee
$\mathcal{S}_m$ being the matter action. The field equations corresponding to the action \eqref{A1} may be expressed as,

\be F'(R)R_{\mu\nu} - {1\over 2}F(R)g_{\mu\nu} - \nabla_\mu\nabla_\nu F'(R) + g_{\mu\nu}\Box F'(R) = \kappa T_{\mu\nu}. \ee
In the above, $\Box F'(R) = {F^{;\mu}}_{;\mu} = \frac{1}{\sqrt{-g}}\partial_\mu(\sqrt{-g}~g^{\mu\nu}\partial_\nu F')$, where the semicolon stands for the covariant derivative, and $\kappa = 8\pi G$. It is noteworthy that $F(R)$ theory of gravity, under weak field approximation leads to the `GTR' \cite{5,6}.\\

However, once flatness is assumed the curvature vanishes and the orientation of vectors remain unaltered under parallel transport along a curve, and the geometry is teleparallel. The condition for torsion based teleparallelism is ${\bar R^\alpha}_{\beta\mu\nu} = 0 = {\tilde R^\alpha}_{\beta\mu\nu}$, where over-bar and over-tilde stand for the Riemann curvature formed from the Weitzenb\"ock connections and connections associated with the non-metricity tensor respectively. The role of $R$ is played by the torsion scalar $\mathrm{T}$ and the action for generalized gravity theory with torsion \cite{7} may be expressed as,

\be \label{EP} A = \int|e| F(\mathrm T)d^4 x +\mathcal{S}_m, \ee
where, $|e|$ = det $e^{i}_{\mu}=\sqrt {-g}$, $e^{i}_{\mu}$ being the components of the vierbien field $\mathbf{e_{i}}(x^\mu)=e^{\mu}_{i}\partial\mu$. The metric tensor is obtained from the dual vierbein as $ g_{\mu\nu}(x)=\eta_{ij}e^{i}_{\mu}(x) e^{j}_{\nu}(x)$. The role of the Ricci scalar ($R$) in the Einstein-Hilbert action is played in metric teleparallel theory by the torsion scalar ($\mathrm {T}$), which is found from the torsion tensor

\be {\mathrm{T}^{\rho}}_{\mu\nu} \equiv {e^\rho}_{i}[\partial_{\mu}{e^i}_{\nu}-\partial_{\nu}{e^i}_{\mu}],\ee
as,
\be\label{TS} \mathrm{T}  = {S_\rho}^{\mu\nu} {\mathrm {T}^\rho}_{\mu\nu},\;\;\mathrm{where},\;\;{S_\rho}^{\mu\nu} = {1\over 2}\left({K_\rho}^{\mu\nu}+\partial^{\mu}_\rho {\mathrm{T}_\alpha}^{\alpha\nu}-\partial^{\nu}_\rho{\mathrm{T}_\alpha}^{\alpha\mu}\right),\ee
is called the super-potential, and the affine connection takes the Weitzenb\"ock form,

\be\label{W} {\Gamma^\alpha}_{\mu\nu} = {e^\alpha}_\lambda \partial_{\nu} {e^\lambda}_\mu.\ee
The above action \eqref{EP} leads to the following field equations for $F(\mathrm T)$ gravity,

\be\label{ft} e^{-1}\partial{\mu}(e S^{\mu\nu}_{i})F_{,\mathrm{T}}-e^{\lambda}_{i}\mathrm{T}^{\rho}_{\mu\lambda}S^{\nu\mu}_{\rho}F_{,\mathrm{T}}
+S^{\mu\nu}_{i}\partial_{}\mu(F(\mathrm{T}))F_{,\mathrm{TT}}+
\frac{1}{4}e^{\nu}_{i}F(\mathrm{T})=\frac{1}{4}e^{\rho}_{i} {T}^{\nu}_{\rho},\ee
where ${S^{\mu\nu}}_{i}={e^\rho}_{i}{S^{\mu\nu}}_{\rho}$ and $F_{,\mathrm{T}}$ denotes differentiation with respect to $\mathrm{T}$.\\

In symmetric teleparallel gravity on the contrary, the role of Ricci scalar is played by the non-metricity scalar ($Q$), and again the generalized action is given by,

\be \label{SP} A = \int F(Q)\sqrt{-g} d^4 x +\mathcal{S}_m, \ee
where,

\be\label{QS}\begin{split}&  Q_{\alpha\mu\nu} = \frac{\partial g_{\mu\nu}}{\partial x^\alpha} - {\Gamma^\beta}_{\alpha\mu}g_{\beta\nu} -  {\Gamma^\beta}_{\alpha\nu}g_{\mu\beta};\\&
Q_\alpha= g^{\mu\nu}Q_{\alpha\mu\nu} = {Q_{\alpha\nu}}^\nu;\;\;\;\;\;\;\;\;\; \hat{Q}_\alpha= g^{\mu\nu}Q_{\mu\alpha\nu} = {Q_{\nu\alpha}}^\nu;\\&
Q = -\frac{1}{4}Q_{\alpha\mu\nu}Q^{\alpha\mu\nu} + \frac{1}{2}Q_{\alpha\mu\nu}Q^{\mu\nu\alpha} + \frac{1}{4}Q_\mu Q^\mu - \frac{1}{2}Q_\mu \hat{Q}^\mu.\end{split}\ee
Varying the above action \eqref{SP} with respect to the metric tensor $g^{\mu\nu}$ and the connection the following field equations are obtained,

\be \label{FE1}
\frac{2}{\sqrt{-g}} \nabla_\lambda (\sqrt{-g}F_Q{P^\lambda}_{\mu\nu}) +\frac{1}{2}F g_{\mu\nu} + F_Q(P_{\nu\rho\sigma} Q_\mu{}^{\rho\sigma}
-2P_{\rho\sigma\mu}Q^{\rho\sigma}{}_\nu) = -\kappa T_{\mu\nu},
\ee
\be\label{Varcon} \nabla_\mu\nabla_\nu\left(\sqrt{-g} F_{,Q} {P^{\mu\nu}}_\lambda\right) = 0,\ee
where,

\be {P^\alpha}_{\mu\nu} = {1\over 2} Q_{(\mu~\nu)}^{~~\alpha} -{1\over 4}{Q^\alpha}_{\mu\nu} +{1\over 4}(Q^\alpha - \hat{Q}^\alpha)g_{\mu\nu}- {1\over 4}{\delta^\alpha}_{(\mu}Q_{\nu)}  \ee
is called the non-metricity conjugate tensor. In the above, we have used first bracket in the suffix to denote symmetrization of indices. The covariant formulation \cite{8} of the field equation \eqref{FE1} is

\be \label{actionqcov}
F_{,Q} {G}_{\mu\nu} + \frac{1}{2}g_{\mu\nu}(Q F_{,Q} - F(Q)) + 2F_{,QQ} {\nabla}_\lambda Q P^\lambda{}_{\mu\nu} = -\kappa T_{\mu\nu}
\ee
where, ${G}_{\mu\nu} = {R}_{\mu\nu} - \frac{1}{2} g_{\mu\nu} {R}$, corresponds to the Levi-Civita connection, and $F_{,Q}$ stands for derivative of $F(Q)$ with respect to $Q$.\\

Having briefly discussed the three different ways attributed to the gravity, sometimes dubbed as the `geometrical trinity of gravity', we now proceed in the following section to find the specific forms of $F(R)$, $F(\mathrm{T})$ and $F(Q)$.

\section{Reconstruction from the radiation era:}

As mentioned, standard FLRW model of cosmology fits best in the radiation and early matter dominated era. In particular, a Friedmann-like radiation dominated era expedites the standard Big-Bang-Nucleosynthesis (BBN), accounts for the observed structure formation (once the seeds of structure are given) and the Cosmic Microwave Background Radiation (CMBR). Therefore, exploring suitable forms of different theories of gravity, in view of early Friedmann-like radiation era is supposed to be the best option. It may be mentioned that the modified or generalized alternative theories of gravity might admit Friedmann-like solutions depending on their forms. For example, the first order Palatini theory of gravity and also the two telleparallel theories viz., TEGR and STEGR are equivalent to GTR. As a result, Friedmann solutions automatically follows. We also remind that if modified gravity includes terms like $F(R) = \alpha R + \beta R^2 + \gamma R^n$ or else, so that $R$ dominates in the middle, then Friedmann-like solutions result in the radiation as well as in the early matter-dominated eras. Nonetheless, modified and alternative theories should be treated as viable only if these theories admit Friedmann-like solutions in the radiation and early matter dominated eras. Therefore, in this article we restrict ourself to the cosmological principle, considering the isotropic and homogeneous Robertson-Walker (RW) space-time,

\be\label{RW} ds^2 = -dt^2 + a^2(t)\left[\frac{dr^2}{1-kr^2} + r^2(d\theta^2 + r^2 sin^2 \theta d\phi^2)\right],\ee
and explore the relevant forms of modified gravity theories comprised with the curvature [$F(R)$], torsion [$F(\mathrm T)$], as well as non-metricity [$F(Q)$].

\subsection{$F(R)$ gravity:}

The Ricci scalar corresponding to the RW metric \eqref{RW} is,

\be \label{RS}  R = 6\left(\frac{\ddot a}{a} + \frac{\dot a^2}{a^2} + \frac{k}{a^2}\right).\ee
The point Lagrangian corresponding to the action \eqref{A1} is given by \cite{9a}

\be L_R = -6a\dot a^2 F' - 6a^2 \dot a\dot R F'' + a^3(F - RF') + 6 kaF' + L_m,\ee
$L_m$ being the matter Lagrangian, comprised of a barotropic fluid associated with the thermodynamic pressure ($p$) and energy density ($\rho$) along with the cold dark matter (CDM). The field equations are,

\be \begin{split}&\left(2{\ddot a\over a} + {\dot a^2\over a^2} + {k\over a^2}\right)F' + \left(\ddot R + 2{\dot a\over a}\dot R\right) F'' +\dot R^2 F''' +{1\over 2}(F - RF') = - p,\\&
\left({\dot a^2\over a^2} + {k\over a^2}\right)F' + {\dot a\over a}\dot R F'' +{1\over 6}(F - RF') = {\rho\over 3},\end{split}\ee
Adding the above pair of equations, one obtains,

\be 2\left({\ddot a\over a} + {\dot a^2\over a^2} + {k\over a^2}\right)F' + \left(3{\dot a\over a}\dot R + \ddot R\right)F'' + +\dot R^2 F''' +{2\over 3}(F - RF') = {\rho\over 3} -p.\ee
Clearly, the radiation era ($p = {1\over 3}\rho$) does not evolve like standard (FLRW) model ($a \propto \sqrt t$), for the spatially flat ($k = 0$) universe, and as such a viable form of $F(R)$ remains obscure. For $k\ne 0$, on the other hand, the differential equation cannot be solved, either in the radiation or in the vacuum dominated ($\rho = 0 = p$) era, other than the fact that, for $k = 0$ vacuum era admits a de-Sitter solution. On the contrary, it may be mentioned that different forms of $F(R)$ emerge in different eras, in view of Noether symmetry \cite{9b}.\\

Nonetheless, interesting results emerge from a generalized four-dimensional string effective action associated with higher order curvature invariant terms, being expressed in the following form \cite{9c},

\be \label{AC} A = \int \left[f(\phi) R+B F(R) - {\omega(\phi)\over \phi}\phi_{,\mu}\phi^{^,\mu} - V(\phi) - \kappa\mathcal{L}_m \right]\sqrt{-g} d^4 x, \ee
where $B$ is the coupling constant, $f(\phi)$ is the coupling parameter, $\omega(\phi)$ is the variable Brans-Dicke parameter and $\mathcal{L}_m$ is the matter Lagrangian density. It had been revealed \cite{9c} that the above action admits a conserved current $J^{\mu}= (3f'^2+2f{\omega\over \phi})^{1\over 2}$ under the condition,

\be\label{R.1} B\left[R F_{,R} +3(F_{,R})^{;\alpha}_ {;\alpha}-2f\right]= {\kappa\over 2}T^\mu_\mu, ~~~\text{provided}~~~ V=\lambda f^2,\ee
where $\lambda$ is a constant. To perceive the very important role of such a conserved current, a particular case was also studied \cite{9c}. For example, considering $F(R) = R^2$, the condition for the existence of conserved current \eqref{R.1} reads as,

\be\label{R.2} {R^{;\mu}}_ {;\mu}=\Box R=\left(\ddot R+3{\dot a\over a}R\right)={\kappa\over 12B} T^\mu_{\mu}.\ee
Now if the scalar field ($\phi$) is completely used up in the process of driving inflation and reheating under particle creation, then since in the  radiation-dominated era ${T^\mu}_\mu = T = \rho - 3p =0$, therefore in view of \eqref{R.2} $\Box R = \left(\ddot R+3{\dot a\over a}R\right) = 0$.
Further, since in the very early vacuum dominated era sufficient inflation makes the universe spatially flat, i.e. $k = 0$, so the above equation \eqref{R.2} admits a Friedmann like solution, $a=a_0\sqrt t$ in the radiation dominated era as demonstrated in \cite{9c}. This is a unique result, since even in the presence of higher order curvature invariant term Friedmann-like radiation era is admissible.\\

Finally, we consider pure curvature induced gravity theory, a yet another generalized form of $F(R)$ action, considered in \cite{10} being expressed as,

\be \label{ABK} A = \int \left[\alpha R+B R^{3\over 2} +\gamma R^2- \mathcal{L}_m \right]\sqrt{-g} d^4 x. \ee
The above action was found more suitable to explain cosmic evolution right from the very early stage till date, since it satisfies all the strong conditions necessary for a viable $F(R)$ theory of gravity. It may be mentioned that $R^{3\over 2}$ term appeared as a consequence of Noether symmetry in R-W metric both in vacuum as well as in the matter dominated eras \cite{{11}, {12}, {13}, {14}, {15}}. At the initial stage, $R^2$ term dominates and a de-Sitter solution is realizable. This leads to the inflationary epoch and reheating following the mechanism of particle production via scalaron decay, exploiting gravity only without invoking phase transition \cite{{16}, {17}}. After the reheating is over, the universe evolves as Friedmann-like radiation $(a\propto \sqrt t)$ and early matter $(a\propto t^{2\over 3})$ dominated eras, and finally accelerated expansion is realized due to the presence of the combination of linear term $R$ and non-linear term $R^{3\over 2}$ \cite{13}. Further, following numerical analysis \cite{10} taking deceleration parameter $q$ as a function of the red-shift parameter $z$, three distinct cases were analyzed to establish the fact that, within a particular range of values for $\beta$ and $\gamma$, $q$ versus $z$ plot depicts that the universe was in pure radiation era at $z > 3200$. Thereafter, deceleration parameter falls off from the matter-radiation equality epoch to the decoupling epoch. It falls even sharply thereafter and a Friedmann type $(q \approx 0.5)$ matter dominated era is reached at around $z \approx 200$. The deceleration parameter then starts increasing slowly and it is peaked to $(q = 1)$, explaining re-ionization of the inter galactic medium (IGM). Subsequently, late time accelerated expansion is initiated and the phantom divide line is crossed, to make a second transition out of it at $z \approx 0.5$.\\

In a nutshell, higher order gravity theory is a viable option to explain cosmological evolution, from the early vacuum dominated era till date via a Friedmann-like radiation and early matter dominated eras. Additionally, re-ionization of the IGM may also be explained in the process. Inflation in the Starobinski model \cite{{16}, {17}} with $R^2$ term is widely explored in the literature, and therefore we leave it here.

\subsection{Torsion-based Metric $F(\mathrm T)$ Teleparallel gravity:}

Although, modified $F(R)$ theory of gravity seems to produce a viable theory of gravity, the fact that the field equations are fourth order,it is vulnerable to Ostrogradsky's instability and also suffers from the violation of the superposition principle. This led to consider alternative telleparallel theories of gravity, already discussed. The main virtue of telleparallel theory of gravity is that the field equations are second order as in the case of Lanczos-Lovelock gravity. Yet another possible fundamental motivation for teleparallelism is that, it was thought to account for the otherwise anomalous dimension and hence non-renormalisability of the GTR action, providing a totally different approach towards a ultra-violet completion of gravity.  Here, we shall consider generalized metric telleparallel gravity theory. \\

In the recent years, a generalized version of the `teleparallel gravity' with torsion, namely the $F(\mathrm T)$ theory of gravity (where, $\mathrm T$ stands for the torsion scalar), also dubbed as generalized `gravity with torsion' has been proposed as an alternative to both the dark energy theories and the modified theories of gravity. Primarily $F(\mathrm T)$ theory of gravity was  proposed to drive inflation. Later, it was applied to drive the current accelerated expansion of the present universe without considering dark energy \cite{18, 19}. It is worth mentioning that `teleparallel equivalent of general relativity' (TEGR) is established for $F(\mathrm T)\propto\mathrm{T}$, since dynamically it leads to GTR. Further as demonstrated in \cite{10}, the trace of electro-magnetic \ radiation  field tensor $(T = \rho-3p)$ being zero, the contracted GTR equation $R \propto{T}$ enforces Ricci scalar to vanish $(R = 0)$ and the Friedmann-like decelerated expansion $[a(t)\propto \sqrt t]$ results in automatically in the radiation dominated era. This is true also in the case of modified theory of gravity, as demonstrated above. However, in the case of torsion, although $F(\mathrm T)\propto \mathrm{T}$ leads to GTR, the trace of the energy-momentum tensor $T = 0$, does not lead to the static solution $\mathrm{T} = 0$. Therefore, even though all the results of GTR hold, the pathology of discontinuous evolution of the Ricci scalar (large initially, vanishing in the middle and small at present) is averted. However unlike GTR and $F(R)$ theories, gravity with torsion is not generally covariant by default. Nonetheless, it has been argued that it may be made so by introducing a new variable, viz, a spin-connection \cite{7,20}. The reason being, the spin-connection enters the teleparallel action only as a surface term and does not contribute to the field equations. \\

Now, the components of vierbein field for the RW metric \eqref{RW} are expressed in terms of the cosmological scale factor $a(t)$ as,

\be\label{2.2}e^{i}_{\mu}= \mathrm{diag}(1,a(t),a(t),a(t)),\ee
and the torsion scalar reads as \cite{21},

\be\label{T} \mathrm {T} = -6H^2+{6k\over a^2},\ee
where, $H = {\dot a\over a}$ is the Hubble parameter. Thus the field equations \eqref{ft} for RW metric are,

\be\label{T1} 12 H^2F_{,\mathrm T} + F(\mathrm T)=\rho,\ee								
\be\label{T2} 4\left({k\over a^2}+{\dot H}\right)\left[12H^2 F_{,\mathrm T\mathrm T}+F_{,\mathrm T}\right] -F({\mathrm T}) - 4F_{,\mathrm T}[2{\dot H}+3H^2]= p,\ee
where $\rho$ and $p$ are the energy density and thermodynamic pressure of a barotropic fluid (inclusive of dark matter component) respectively and $F_{,\mathrm T}$ stands for derivative of $F(\mathrm T)$ with respect to $\mathrm T$. Bianchi identity does not hold naturally in teleparallel gravity theories. However, in the RW metric under consideration, it holds even for extended models such as $F({\mathrm T})$ \cite{21}. For example, taking the time derivative of the first equation \eqref{T1}, and also adding the the two \eqref{T1} and \eqref{T2}, one gets,

\be\begin{split} &{\dot \rho}=12H^2\dot {\mathrm T} F_{,\mathrm T \mathrm T}+24 H\dot{ H }F_{,\mathrm T}+\dot {\mathrm T} F_{,\mathrm T}=-144H^3{\dot H}F_{,\mathrm T \mathrm T}-\frac{144 kH^3F_{,\mathrm T \mathrm T}}{a^2}+12H{\dot H}F_{,\mathrm T}-\frac{12kHF_{,\mathrm T}}{a^2},~\text{and}\\&\left(\rho+p\right)=48{\dot H}H^2 F_{,\mathrm T\mathrm T}-4\dot H F_{,\mathrm T}+{48k\over a^2}H^2 F_{,\mathrm T\mathrm T}+{4k\over a^2} F_{,\mathrm T}.\end{split}\ee
Combining the above two one finds

\be\label{BI} \dot \rho + 3H(\rho+p) = 0,\ee
which is the energy-momentum conservation law, an outcome Bianchi identity. Thus, one may also write

\be\label{BI1} \rho = \rho_0 a^{-3(1+\omega)},\ee
for the barotropic equation of state $p = \omega \rho$, as usual, where $\omega$ is the state parameter.\\

Let us now proceed to find the form of $F(\mathrm{T})$ for flat space $(k=0)$, already obtained earlier in view of cosmological evolution \cite{22}. First, combining equations \eqref{T1} and \eqref{T2} for flat space $k=0$, one finds for the vacuum era ($p = 0 = \rho$),

\be\label{CT} \dot H\left[12 H^2F_{,\mathrm T \mathrm T} - F_{,\mathrm T}\right]=0.\ee
Clearly, two possibilities emerge from the equation \eqref{CT}: i) $F = F_0\sqrt {\mathrm T}$, where $F_0$ is a constant of integration. In view of this form of $F(\mathrm{T})$, either of the field equations \eqref{T1} or \eqref{T2} yields simply the definition and hence, no dynamics results. Indeed it is expected, since the above form of $F(\mathrm{T})$ only results in a divergent term in the action. ii) $H = \lambda$ (where $\lambda$ is a constant) leading to the de-Sitter solution $a = a_0\exp({\lambda t})$, while $F(\mathrm T) = F_0 \exp(-{t\over 12\Lambda^2})$ emerges as an exponentially decaying function, which is not much promising.\\

Since vacuum dominated era does not yield a reasonably viable form of  $F(\mathrm{T})$, so let us now advance further to study the radiation-dominated era. Starting from the action \eqref{EP}, if a solution in the form $a = a_0t^n$ is sought in the radiation dominated era ($p ={1\over 3}\rho$), for which the energy-momentum conservation law yields $\rho a^4 = \rho_{r0}$, where $a_0$, $n$ and $\rho_{r0}$ are constants, then the following form of $F(\mathrm T)$ is found \cite{9b},

\be\label{FT} F(\mathrm T)_r = 2f_1\sqrt{-6\mathrm {T}}+\frac{\rho_{r0}}{6a^4_0(4n-1)n^{4n}(-6)^{2n-1}}{\mathrm{T}}^{2n}, \ee
where $f_1$ is a constant and the suffix $r$ stands for radiation dominated era. Note that the first term of the above equation is essentially a divergent term in the RW metric under consideration. Thus, only the second term is left and hence,

\be\label{FT1} F(\mathrm T)_r = \frac{\rho_{r0}}{6a^4_0(4n-1)n^{4n}(-6)^{2n-1}}{\mathrm{T}}^{2n}. \ee
It is quite apparent that under the choice $n = {1\over 2}$, the radiation era evolves exactly like the standard (FLRW) model, however the action too reduces to that of GTR  since $[F(T)\propto \mathrm T]$ (apart from a total derivative term), which is TEGR, as already mentioned. Of-course, $n = {3\over 4}$ also leads to decelerated expansion, but such a slow deceleration tells upon the formation of CMBR at latter times than observed. Let us next, focus on the matter dominated era ($p = 0$), for which the energy-momentum conservation law yields $\rho a^3 = \rho_{m0}$, (where $ \rho_{m0}$ is a constant, and suffix $m$ stands for the matter dominated era). Proceeding in the similar manner as before, one finds,

\be \label{ftM} F(\mathrm T)_m ={{\rho_{m0}} \over 3a_0^3(3n-1)n^{3n}(-6)^{{3n\over 2}-1}}{\mathrm{T}}^{3n\over 2},\ee
where the suffix $m$ stands for the matter dominated era. Thus, for $n = {2\over 3}$,  the matter dominated era evolves ($a \propto  t^{2\over 3}$) in the same manner as the standard (FLRW) model of cosmology, and GTR is recovered through TEGR. Whatsoever, it was also discussed in \cite{9b} that when torsion is attributed to gravity, usually a form such as $F(\mathrm T ) = f_0\mathrm{T} + f_1{\mathrm T}^2+\cdots$ is chosen to combat early deceleration in the Friedmann form $[a(t) \propto t^{2\over 3}]$ followed by late-time cosmic acceleration in the matter (pressure-less dust) dominated era. In view of \eqref{FT} and \eqref{ftM}, it is clear that $F(\mathrm{T})\propto \mathrm T$ gives exactly Friedmann-like radiation dominated era $[a(t) \propto t^{1\over 2}]$, and early pressure-less dust dominated era $[a(t) \propto t^{2\over 3}]$ respectively. Thus, following generalized form of $F(\mathrm{T})$,

\be F(\mathrm T) = f_0 \mathrm T + f_1 (-\mathrm T)^{3\over 2} + f_2 ({\mathrm T}^2) + \cdots\ee
might be useful to study cosmic evolution. On the contrary, Noether symmetry analysis \cite{9b} demands that instead of $\mathrm T^2$, one should associate $\mathrm T^3$ and higher odd integral powers in the action. That is, a viable form that might explain the cosmic evolutionary history may be in the form, $F(\mathrm T ) = f_0\mathrm{T} + f_1{\mathrm T}^3+\cdots$. It may be mentioned that the current analysis also reveals the fact that pure $F(\mathrm T $) gravity in vacuum ($\rho = p = 0$) does not give rise to any dynamics. Therefore to drive inflation, and also to avert the pathological behaviour of pure $F(\mathrm T )$ gravity in the very early vacuum-dominated era, either unimodular $F(\mathrm T)$ gravity has to be considered \cite{23} or a scalar field should be associated with $F(\mathrm T )$ gravity theory \cite{22} and the action may be proposed as,

\be\label{A2} A= \int\left[f_0\mathrm{T} + f_1{\mathrm T}^3+\cdots-{1\over 2}\phi_{,\mu}\phi^{,\mu}-V(\phi)\right]\sqrt{-g}d^4x,  \ee
where, the scalar field drives inflation at the very early stage and decays to an insignificant value in the process of particle creation. Therefore once the inflation is over, the universe enters the radiation and thereafter the matter dominated eras, whence $\mathrm T$ dominates to envisage the standard model. Now since $\mathrm T = - 6 H^2 < 0$ by definition, therefore as the Hubble parameter decreases further, the odd-integral higher degree terms start dominating and become responsible for late-time cosmic acceleration. The great conceptual advantage over modified theories of gravity is that, unlike the Ricci scalar, $\mathrm T \ne 0$, at any stage of cosmic evolution in the middle.

\subsubsection{Energy conditions:}

Before we proceed further, it is necessary to fix the signature of the coefficients $(f_0, f_1\mathrm{and}~ f_2 )$ associated with the two different forms of $F(\mathrm T)$ obtained above, viz., i) $F(\mathrm T)=f_0 \mathrm T + f_1 (-\mathrm T)^{3\over 2} + f_2 ({\mathrm T}^2) + \cdots$ and ii) $F(\mathrm T ) = f_0\mathrm{T} + f_1{\mathrm T}^3+f_2{\mathrm T}^5+\cdots$, in view of the energy conditions. For perfect fluid $T_{\mu\nu} = (\rho + p)u_\mu u_\nu + p g_{\mu\nu}$, the energy conditions are:\\
1. Null energy condition: $\rho + p \ge 0$.\\
2. Weak energy condition: $\rho + p \ge 0$ and $\rho \ge 0$.\\
3. Dominant energy condition:$\rho \ge |p|$.\\
4. Strong energy condition:$\rho + p \ge 0$ and $\rho + 3p \ge 0$.\\
Note that, there is no restriction on the thermodynamic pressure, i.e., $p < 0$ is allowed, nonetheless, if both $\rho > 0$, and $p > 0$, all energy conditions are satisfied simultaneously.\\

\noindent
\textbf{Form-1: Fixing coefficients of $F(\mathrm T)=f_0 \mathrm T + f_1 (-\mathrm T)^{3\over 2} + f_2 ({\mathrm T}^2) + \cdots$.}\\

\noindent
To expatiate the signature associated with the coefficients of $F(\mathrm{T})$, let us analyse the form of for each individual term separately.

\noindent
1. For, $F(\mathrm T)=f_0 \mathrm T $, the field equations \eqref{T1} and \eqref{T2} read as,

\be 6f_0H^2=\rho,~~~\text{and}~~~ -6f_0 H^2-4f_0{\dot H} = p ,\ee respectively.
Therefore, $f_0$ has to be positive so that $\rho >0$. Consequently, $\rho+p =-4f_0{\dot H} >0$ is also satisfied, because $\dot H<0$, in the expanding model.\\

\noindent
2. Again for, $F(\mathrm T)=f_1(-\mathrm T)^{3\over 2}$, the field equation \eqref{T1} are,

\be 24f_1\sqrt6 H^3=\rho, ~~~\text{and}~~~-\frac{9}{2}f_1\sqrt6{\dot H}H-24f_1\sqrt6 H^3=p, \ee
so to keep $\rho$ positive, $f_1$ has to be positive and $\rho+p =- \frac{9}{2}f_1\sqrt6{\dot H}H >0$ is also satisfied since, $\dot H<0$.\\

\noindent
3. Finally for, $F(\mathrm T)=f_2 {\mathrm T}^2 $, the field equations \eqref{T1} and \eqref{T2} may be cast as,

\be -108f_2H^4=\rho,~~~\text{and} ~~~144f_2{\dot H}H^2+108f_2 H^4=p,\ee
respectively. Therefore, $\rho > 0$ is ensured provided $f_2 <0$ and finally to satisfy weak energy condition, $\rho+p =144 f_2{\dot H}H^2>0$ also ensures $f_2 <0$, since $\dot H < 0$, in the expanding model.\\

\noindent
Thus all the energy conditions are satisfied provided $F(T)$ has the form

\be\label{ft1}F({\mathrm T})=f_0 \mathrm T + f_1 (-\mathrm T)^{3\over 2} - f_2 ({\mathrm T}^2),\ee
in which all the coefficients are positive, i.e., $f_0 > 0$, $f_1 > 0$, and $f_2 > 0$, and we terminate after the third term\\

\noindent
\textbf{Form-2: Fixing coefficients of $F(\mathrm T)=f_0 \mathrm T + f_1 (\mathrm T^3) + f_2 ({\mathrm T}^5) + \cdots$.}\\

\noindent
Likewise, let us fix the signature of the coefficients ($f_0$, $f_1$ and $f_2$) appearing in the second form of $F(\mathrm{T})$, considering each term separately.

\noindent
1. For, $F(\mathrm T)=f_0 \mathrm T $, the field equations \eqref{T1} and \eqref{T2} are expressed as,

\be 6f_0H^2=\rho,~~~\text{and}~~~ -6f_0 H^2-4f_0{\dot H} =p ,\ee respectively.
Therefore, to ensure $\rho >0$, $f_0$ must be positive and thus $\rho+p =-4f_0{\dot H} > 0$ is also satisfied as $\dot H<0$ in an expanding model.\\

\noindent
2. Again for, $F(\mathrm T)=f_1(\mathrm T)^{3}$, the field equation \eqref{T1}and \eqref{T2} are,

\be  1512 f_1H^6=\rho,~~~\text{and}~~~-2160f_1{\dot H}H^4-1512f_1H^6=p.\ee
Therefore, $\rho >0$ is ensured provided $f_1 >0$ and also $\rho+p =-2160 f_1{\dot H}H^4>0$ is ensures if $f_1 >0$ as $\dot H<0$, and as a result the weak energy condition is also satisfied.\\

\noindent
3. Finally for, $F(\mathrm T)=f_2 {\mathrm T}^5 $, the field equation \eqref{T1} take the form,

\be 69984f_2H^{10}=\rho,~~~\text{and}~~~181440f_2{\dot H} H^8- 69984f_2H^{10}=p.\ee
Hence, $\rho > 0$ together with $\rho+p = 181440f_2{\dot H}H^8 > 0$, provided $f_2 > 0$ as $\dot H < 0$ and so the weak energy condition is also satisfied.\\

\noindent
Thus, all the energy condition are satisfied, provided the form of $F({\mathrm T})$ is

\be \label{ft2}F({\mathrm T})=f_0 \mathrm T + f_1 (\mathrm T)^{3} + f_2 ({\mathrm T}^5),\ee
where, all the coefficients $f_0, f_1, f_2$ are positive, considering no additional terms.

\subsubsection{Slow roll Inflation:}

We have seen that both the forms $F(\mathrm{T})$ of generalized metric teleparallel gravity theory presented in \eqref{ft1} and \eqref{ft2} admit FLRW type radiation and early matter dominated eras, and can trigger accelerated expansion in the late stage of cosmological evolution. Further, both the forms are validated by energy conditions. To explore their behaviour in the very early universe, let us consider slow roll inflation for both the models to find the inflationary parameters and to compare with the currently observed constraints.\\

Cosmological inflation, that occurred sometimes between $10^{-42}~s$ and $10^{-32\pm6}~s$, not only can solve the horizon, flatness and monopole problems but also generates the seeds of perturbation required to trigger the structure formation at a latter epoch. Although, it is a quantum theory of perturbation, where gravity is treated as classical, while all other fields remain quantized, classical field equations are well suited to study inflation. The recently released data sets \cite{24,25} imposed tighter constraint on  the inflationary parameters $n_s$ ($0.9631 \le n_s \le 0.9705$), as well as on the tensor to scalar ratio ($r < 0.055$). More recently, combination of Planck PR4 data with ground-based experiments such as, BICEP/Keck 2018 (BK18), BAO and CMB lensing data, tightens the tensor to scalar ratio even further to $r < 0.032$ \cite{26}. Nonetheless in recent years $r$ has been constrained starting from $r < 0.14$ to the above mentioned value and therefore we presume that $r$ might be restricted to even less value in more precise future experiments, such as polarized CMB space missions (including LiteBIRD and OKEANOS) \cite{27}. To study inflation, we incorporate a scalar field $\phi$ along with a potential $V(\phi)$ in the action, which drives the inflation, as already mentioned in \eqref{A2}. As an example, let us choose a special form of potential $V = V_0-{V_1\over \phi}$, such that when $\phi$ becomes large enough, then $V \approx V_0$, representing a flat potential. \\

\noindent
\textbf{Form-1: $F(\mathrm T)=f_0 \mathrm T + f_1 (-\mathrm T)^{3\over 2} -f_2 ({\mathrm T}^2) + \cdots$,}\\

\noindent
As the energy condition ensures all the coefficients are positive, so in the vacuum era, the field equation \eqref{T1}, and the $\phi$ variation equation are expressed as,

\be\label{E1} 6f_0 H^2+24{\sqrt 6}f_1 H^3+ 108 f_2H^4= V(\phi)+ {1\over 2}\dot\phi^2, \hspace{1cm}\ddot{\phi} +3H|\dot\phi|+V'=0. \ee
Let us now consider the standard slow-roll conditions $\dot\phi^2\ll V(\phi)$ and $|\ddot\phi|\ll 3{H}|\dot\phi|$, on the pair of equations (\ref{E1}), which therefore finally reduce to,

\be\label{com1} 108f_2 H^4+24\sqrt{6}f_1H^3+ 6f_0 H^2- V(\phi)=0, \hspace{0.7cm} 3H\dot\phi+V'=0.\ee
Clearly, the first quartic equation \eqref{com1} has four roots, each of which are exorbitantly large and complicated including cubic roots. Hence, these three terms together in $F(T)$ cannot be handled. Therefore in the following, we consider the combination of (i) the first term ($\mathrm T$) and the third term ($\mathrm T^2$) and (ii) the first two terms i.e., $\mathrm T$ and $({-\mathrm T})^{3\over 2}$.\\

\noindent
\textbf{Case-1: $F(\mathrm T)=f_0 \mathrm T - f_2 {\mathrm T}^2$:}\\

\noindent
Earlier, slow roll inflation considering the form $F(\mathrm T)=f_0 \mathrm T + f_2 {\mathrm T}^2$, where both $f_0 > 0$ and $f_2 > 0$ has been studied extensively and wonderful agreement with the observed data was found \cite{22}. Nonetheless, the energy condition suggests that ${\mathrm T}^2$ term must appear with a negative sign. Hence it is required to look over if the inflationary parameters are still at par with the observational data. The field equations are now,

\be 6f_0 H^2+108f_2 H^4={1\over 2}{\dot \phi}^2+V(\phi),\hspace{1cm}\ddot{\phi} +3H|\dot\phi|+V'(\phi)=0,\ee
upon which we apply the standard slow-roll conditions $\dot\phi^2\ll V(\phi)$ and $|\ddot\phi|\ll 3{H}|\dot\phi|$ to finally obtain,

\be\label{T21} \gamma H^4+ 6f_0 H^2- V(\phi)=0, ~~~~~~~ 3H\dot\phi+V'(\phi)=0,\ee
where $\gamma= 108f_2$. Now for the above choice of the potential $V = V_0-{V_1\over \phi}$, the slow roll parameters are expressed as,

\be\label{para2}\begin {split} &\epsilon =  \frac{\gamma^2 {V_1}^2}{12\phi^4\left(\sqrt{9f_0^2+\gamma V_1\left(\frac{V_0}{V_1}-\frac{1}{\phi}\right)}\right)\left[-3f_0 + \sqrt{9f_0^2+\gamma V_1\left(\frac{V_0}{V_1}-\frac{1}{\phi}\right)}\right]^2};\hspace{0.3 in}\eta =-\frac{2M_P^2}{\phi^2(\frac{V_0}{V_1}\phi-1)},\\& N=\int_{\phi_f}^{\phi_i}\frac{3\phi^2\left[-3f_0 +\sqrt{9f_0^2+\gamma V_1\left(\frac{V_0}{V_1}-\frac{1}{\phi}\right)}\right]}{\gamma V_1}d\phi,\end{split}\ee
where $N$ is the number of e-folds. In yet another paper (under preparation) we have found excellent agreement of the inflationary parameters with the observational data. While $N = 50$ is sufficient to solve the horizon and flatness problems, the oscillatory behaviour of the scalar field has also been explored. As an example, we present a set of data in the following Table-1.
\begin{table}
  \begin{minipage}[h]{0.97\textwidth}
  \centering
      \begin{tabular}{|c|c|c|c|c|c|c|c|}
  \hline\hline
      $\phi_i~M_P$ &${V_0\over V_1}~ M_P^{-1}$ & $\gamma V_1~ M_P^{5}$& $f_0~M_P^2$&$\phi_f~M_P$&$n_s$ & $r$ & ${N}$ \\
      \hline
        3.0 &  6.5 & 0.1&0.5 &0.69599&0.96425&0.03128&50 \\
       \hline\hline
    \end{tabular}
     \captionof{table}{Data set for the inflationary parameters for $F(\mathrm T)=f_0 \mathrm T - f_2 {\mathrm T}^2$.}
      \label{tab:Table1}
      \end{minipage}
  \end{table}
It is also important to mention that one may fix the value of $\gamma$, which is arbitrary, to set the energy scale of inflation $H_*$ at the sub-Planckian scale.\\

\noindent
\textbf{Case-2: $F(\mathrm T)=f_0 \mathrm T + f_1 (-\mathrm T)^{3\over 2}$:}\\

\noindent
This case has never been studied earlier. For the above form of $F(\mathrm T)$, we apply the standard slow-roll conditions $\dot\phi^2\ll V(\phi)$ and $|\ddot\phi|\ll 3{H}|\dot\phi|$ and as such the field equations reduce to,

\be \label{T3.2}6f_0 H^2+24{\sqrt 6}f_1 H^3- V(\phi)=0, \hspace{0.5cm} 3H\dot\phi+V'=0.\ee
Now, the first cubic equation of \eqref{T3.2} can be solved for $H$ as,

\be\label{sol} H= {1\over \gamma}\left[-2f_0+\frac{6.35f_0^2 +\Big(V(\phi)\gamma^2-16f_0^3+\sqrt{V(\phi)^2\gamma^4-32V(\phi)\gamma^2f_0^3}\Big)^{2\over 3}}{1.26\big(V\gamma^2-16f_0^3+\sqrt{V(\phi)^2\gamma^4-32V(\phi)\gamma^2f_0^3}\big)^{1\over 3}}\right],\ee
where $\gamma=24{\sqrt 6}f_1 $.  Combining the pair of equations (\ref{T3.2}), one can identify the `potential slow roll parameter' $\epsilon$ with the `Hubble slow roll parameter' ($\epsilon_1$) and also can express $\eta$ and the number of e-folds respectively as,

\be\begin{split}&\epsilon \equiv  -{\dot {H}\over {H}^2}=\frac{V'(\phi) ^2}{3H^4[12f_0+3\gamma H]}, \hspace{1cm} \eta=2 f_0 \left[{V''(\phi)\over V(\phi)}\right]\\& {N}(\phi)\simeq \int_{t_i}^{t_f}{H}dt=\int_{\phi_i}^{\phi_f}{{H}\over {\dot\phi}}d\phi= \int_{\phi_f}^{\phi_i}3\frac{\left[-2f_0+\frac{6.35f_0^2+\Big(V(\phi)\gamma^2-16f_0^3+\sqrt{V(\phi)^2\gamma^4-32V(\phi)\gamma^2f_0^3}\Big)^{2\over 3}}{1.26\big(V\gamma^2-16f_0^3+\sqrt{V(\phi)^2\gamma^4-32V(\phi)\gamma^2f_0^3}\big)^{1\over3}}\right]^2}{\left[\gamma^2V'\right]}d\phi.\end{split}\ee
Now for this potential $V=V_0-{V_1\over \phi}$ under consideration, the expression for $\epsilon$ takes a extortionate form such as,

\be\label{ep}\begin{tiny}\begin{split}\epsilon=&\frac{3.18 V_1^2\gamma^4\Bigg[V_1\gamma^2({V_0\over V_1}-{1\over \phi})-16f_0^3+\sqrt{V_1^2\big({V_0\over V_1}-{1\over \phi}\big)^2\gamma^4-32V_1({V_0\over V_1}-{1\over \phi})\gamma^2f_0^3}\Bigg]^{5\over 3}}{9\phi^4\Bigg[-2.52f_0\big[V_1\gamma^2({V_0\over V_1}-{1\over \phi})-16f_0^3+\sqrt{V_1^2\big({V_0\over V_1}-{1\over \phi}\big)^2\gamma^4-32V_1({V_0\over V_1}-{1\over \phi})\gamma^2f_0^3}\big]^{1\over 3}+6.35 f_0^2+\big[V_1\gamma^2({V_0\over V_1}-{1\over \phi})-16f_0^3+\sqrt{V_1^2\big({V_0\over V_1}-{1\over \phi}\big)^2\gamma^4-32V_1({V_0\over V_1}-{1\over \phi})\gamma^2f_0^3}\big]^{2\over 3}\Bigg]^4}\\&\times\frac{1}{\left[2.52f_0\big[V_1\gamma^2({V_0\over V_1}-{1\over \phi})-16f_0^3+\sqrt{V_1^2\big({V_0\over V_1}-{1\over \phi}\big)^2\gamma^4-32V_1({V_0\over V_1}-{1\over \phi})\gamma^2f_0^3}\big]^{1\over 3}+6.35 f_0^2+\big[V_1\gamma^2({V_0\over V_1}-{1\over \phi})-16f_0^3+\sqrt{V_1^2\big({V_0\over V_1}-{1\over \phi}\big)^2\gamma^4-32V_1({V_0\over V_1}-{1\over \phi})\gamma^2f_0^3}\big]^{2\over 3}\right]},\end{split} \end{tiny} \ee
while, $\eta$ and $N$ may be expressed respectively as,

\be\label{N}\begin{split} &\eta=-\frac{4f_0}{\phi^2(\frac{V_0}{V_1}\phi-1)};\hspace{1.0 cm}
{N}(\phi)\simeq \int_{t_i}^{t_f}{H}dt=\int_{\phi_i}^{\phi_f}{{H}\over {\dot\phi}}d\phi\\&= \int_{\phi_f}^{\phi_i}\Bigg[\frac{3 \phi ^2}{1.588 V_1 \gamma ^2 \left(\sqrt{V_1^2 \gamma ^4 \left(\frac{V_0}{V_1}-\frac{1}{\phi }\right)^2-32 V_1 \gamma ^2 f_0^3 \left(\frac{V_0}{V_1}-\frac{1}{\phi }\right)}+V_1 \gamma ^2 \left(\frac{V_0}{V_1}-\frac{1}{\phi }\right)-16 f^3\right)^{2/3}} \times\\&\Bigg\{-2.52 f_0 \Bigg[{\sqrt{V_1^2 \gamma ^4 \Bigg(\frac{V_0}{V_1}-\frac{1}{\phi }\Bigg)^2-32 V_1 \gamma ^2 f_0^3 \left(\frac{V_0}{V_1}-\frac{1}{\phi }\right)}+V_1 \gamma ^2 \Bigg(\frac{V_0}{V_1}-\frac{1}{\phi }\Bigg)-16 f_0^3}\Bigg]^{1\over 3} \\&+\Bigg(\sqrt{V_1^2 \gamma ^4 \left(\frac{V_0}{V_1}-\frac{1}{\phi }\right)^2-32 V_1 \gamma ^2 f_0^3 \left(\frac{V_0}{V_1}-\frac{1}{\phi }\right)}+V_1 \gamma ^2 \left(\frac{V_0}{V_1}-\frac{1}{\phi }\right)-16 f_0^3\Bigg)^{2/3}+6.35 f_0^2\Bigg\}^2\Bigg]d\phi.\end{split}\ee
Despite such huge structures of the parameters, it is still possible to handle these expressions and we present a table of data-set for the  expressions \eqref{ep} and \eqref{N} in Table-2. In this table, we have varied $\phi_i$ within the range $2.80 ~M_P\leq \phi_i\leq 3.40 ~M_P$, so that $r$ and $n_s$ lie more-or-less within the experimental limit. However, restrictions on $n_s$ and $r$, restricts the number of e-folds within the range $24 \leq N \leq 44$, which still might solve the horizon and flatness problems.\\

\noindent
Let us now compute the energy scale of inflation in view of the relation \eqref{sol}, considering the data: ($N = 40$, for which $\phi_i = 3.3~M_P,~ f_0=0.5~M_p^2, ~{{V_0\over V_1}=6.0 ~M_P^{-1}},~ V_1\gamma^2= 3.4 ~M_P^7$), as depicted in Table-2. Correspondingly we find,

\be\label{ES}{H_*} ={8.114\over \gamma}.\ee
Now, the energy scale of inflation in a single scalar field model in GTR \cite{28} is given by the following expression,

\be\label{wands}{H_*} = 8\times 10^{13}\sqrt{\frac{r}{0.2}}~GeV = 0.96\times 10^{13}~GeV \approx 3.918\times 10^{-6}~M_P,\ee
whose numerical value is computed taking into account the value of the tensor-to-scalar ratio $r = 0.00288$ from the data set of Table-2. Thus, in order to match the scale of inflation \eqref{ES} with the single field scale of inflation \eqref{wands} we are required to constrain $\gamma$, such as $\gamma \approx 2.074 \times 10^{6}~M_P$. Requirement of the sub-Planckian  scale for inflation is the physical ground upon which the parameter $\gamma$ has been constrained and consequently, the values of $V_1$ and $V_0$ are fixed as well,

\be \label{V}V_1= 7.90\times 10^{-11}~M_P^5, \hspace{0.5 in} V_0 = 6 V_1\approx 4.74\times 10^{-10}~M_P^4.\ee

\noindent
Finally, to handle the issue of graceful exit from inflation, we recall the first equation of \eqref{T3.2}, which in view of the above form of the potential, $V(\phi)=V_0-{V_1\over \phi}$, is expressed as,

\begin{align}-\frac{\gamma H^3}{V_1} - \frac{6f_0 H^2}{V_1}+\left[{\dot\phi^2\over 2V_1}+\left({V_0\over V_1}-{1\over {\phi}}\right)\right]=0.\end{align}
During inflation, $H^2$ and $V_1$ are of the same order of magnitude, while the Hubble parameter varies slowly. But, at the end of inflation, the Hubble rate usually decreases sharply and $\gamma H^3$ falls much below $V_1$. Hence, one can neglect both the terms ${\gamma H^3\over V_1}$ and ${f_0 H^2\over V_1}$ without any loss of generality. In the process one finds,

\be{\dot{\phi}}^2=-2\left[V_0-{V_1\over \phi}\right].\label{osc1}\ee
Taking into account, $\phi_f=0.460794 ~M_P$, $V_0= 4.74\times 10^{-10}~M_P^4$ and $V_1= 7.90\times 10^{-11}~M_P^5$ from computation, the above equation clearly exhibits oscillatory behavior of the scalar field $\phi$,

\be\phi=\exp({i\omega t}),\ee
provided, $\omega \approx 5.335\times10^{-5}~M_P$.\\

\begin{figure}
\begin{minipage}[h]{0.97\textwidth}
      \centering
      \begin{tabular}{|c|c|c|c|}
     \hline\hline
      $\phi_i$ in $M_P$  & $n_s$ & $r$ & $ {N}$ \\
      \hline
        2.80 &  0.9656 & 0.00566 &24 \\
        2.85 &  0.9674 &0.00520 &26 \\
        2.90 &   0.9692 & 0.00490 &27 \\
        2.95 &   0.9707 & 0.00456 &28 \\
        3.00 &   0.9723 & 0.00426 &30 \\
        3.10 &   0.9750 & 0.00372 &33\\
        3.20 &   0.9773 & 0.00327 &37 \\
        3.30 &   0.9794 & 0.00288 & 40\\
        3.40 &   0.9812 & 0.00255 &44 \\
\hline\hline
    \end{tabular}
     \captionof{table}{Data set for the inflationary parameters in the case $F(\mathrm T)=f_0 \mathrm T + f_1 (-\mathrm T)^{3\over 2}$, taking into account ${V_0\over V_1}=6 ~M_P^{-1}, ~~V_1\gamma^2= 3.4~M_P^{7}$, ~$f_0=0.5 ~M_P^2$ and varying $\phi$ within the range $2.8 ~M_P< \phi_i < 3.4~M_P$, so that $\phi_f\approx 0.460794~M_P$ at the end of inflation.}
      \label{tab:Table2}
   \end{minipage}
   \end {figure}
In a nutshell, although, we have not been able to handle all the three terms together, nonetheless both the pairs exhibit excellent agreement with the observational data.\\

\noindent
\textbf{Form-2: $F(\mathrm T)=f_0 \mathrm T + f_1 (\mathrm T)^{3} + f_2 ({\mathrm T}^5) + \cdots$,}\\

\noindent
\textbf{Case-3: $F(\mathrm T)=f_0 \mathrm T + f_1 (\mathrm T)^{3}$:}\\
Here too it is impossible to handle all the three terms together, we therefore consider only the first two terms $F(\mathrm T)=f_0 \mathrm T + f_1 (\mathrm T)^{3}$ and focus to study the slow roll inflation. The field equation \eqref{T1} and the $\phi$ variation equation are now expressed as,

\be\label{E2} 6f_0 H^2+ 1080 f_1H^6= V(\phi)+ {1\over 2}\dot\phi^2, \hspace{1cm}
 \ddot{\phi} +3H|\dot\phi|+V'=0.\ee
Applying the standard slow-roll conditions $\dot\phi^2\ll V(\phi)$ and $|\ddot\phi|\ll 3{H}|\dot\phi|$, the pair of equations (\ref{E2}) finally reduce to,
\be\label{com2} \gamma H^6+ 6f_0 H^2- V=0, ~~~~~~~ 3H\dot\phi+V'=0,\ee
where $1080 f_1=\gamma$. Solving for $H^2$ in view of \eqref{com2}, we readily obtain,

\be \label{soln1} H^2 = \frac{\left[\left(V\gamma^2+\sqrt{V^2\gamma^4+32\gamma^3 f_0^3}\right)^{2\over 3}-3.175\gamma f_0\right]}{1.26\gamma\left(V\gamma^2+\sqrt{V^2\gamma^4+32\gamma^3 f_0^3}\right)^{1\over 3}}.\ee
Further, combining equations (\ref{com2}), one can  show that the `potential slow roll parameter' ($\epsilon$) is equal to the `Hubble slow roll parameter' ($\epsilon_1$) under the condition,

\be\label{soln2}\begin{split}&\epsilon \equiv - {\dot {H}\over {H}^2}=\frac{2.52\gamma^4V'^2\left(V\gamma^2+\sqrt{V^2\gamma^4+32\gamma^3 f_0^3}\right)^{4\over 3}}{18\left[\left(V\gamma^2+\sqrt{V^2\gamma^4+32\gamma^3 f_0^3}\right)^{2\over 3}-3.175\gamma f_0\right]^2}\\& \times\frac{1}{\left[\gamma\left[\left(V\gamma^2+\sqrt{V^2\gamma^4+32\gamma^3 f_0^3}\right)^{2\over 3}-3.175\gamma f_0\right]^2+3.175\gamma^2 f_0\left(V\gamma^2+\sqrt{V^2\gamma^4+32\gamma^3 f_0^3}\right)^{2\over 3}\right]};&\\\hspace{0.5in}
\eta = 2 f_0 \left({V''(\phi)\over V(\phi)}\right)\end{split}\ee
Further, one can compute the number of e-folds as,

\be\label{Nphi} {N}(\phi)\simeq \int_{t_i}^{t_f}{H}dt=\int_{\phi_i}^{\phi_f}{{H}\over {\dot\phi}}d\phi= \int_{\phi_f}^{\phi_i}3\frac{\left[\left(V\gamma^2+\sqrt{V^2\gamma^4+32\gamma^3 f_0^3}\right)^{2\over 3}-3.175\gamma f_0\right]}{\left[1.26\gamma V'\left(V\gamma^2+\sqrt{V^2\gamma^4+32\gamma^3 f_0^3}\right)^{1\over 3}\right]}d\phi.\ee
Now under the same choice of the potential as above, viz., $V(\phi)= V_0-{V_1\over \phi}$, the expressions of $\epsilon, ~\eta$ \eqref{soln2} and $N$ \eqref{Nphi} are found as,

\be\label{epsaN}\begin{small}\begin{split}&\epsilon =\frac{2.52\gamma^4V_1^2\left(V_1\gamma^2({V_0\over V_1}-{1\over \phi})+\sqrt{V_1^2\gamma^4({V_0\over V_1}-{1\over \phi})^2+32\gamma^3 f_0^3}\right)^{4\over 3}}{18\phi^2\left[\left(V_1\gamma^2({V_0\over V_1}-{1\over \phi})+\sqrt{V_1^2\gamma^4({V_0\over V_1}-{1\over \phi})^2+32\gamma^3 f_0^3}\right)^{2\over 3}-3.175\gamma f_0\right]^2} \\& \times\frac{1}{\left[\gamma\Bigg[\left(V_1\gamma^2({V_0\over V_1}-{1\over \phi})+\sqrt{V_1^2\gamma^4({V_0\over V_1}-{1\over \phi})^2+32\gamma^3 f_0^3}\right)^{2\over 3}-3.175\gamma f_0\Bigg]^2+3.175\gamma^2 f_0\Big\{V_1\gamma^2({V_0\over V_1}-{1\over \phi})+\sqrt{V_1^2\gamma^4({V_0\over V_1}-{1\over \phi})^2+32\gamma^3 f_0^3}\Big\}^{2\over 3}\right]}\\&
 \eta =-\frac{4f_0}{\phi^2(\frac{V_0}{V_1}\phi-1)},\hspace{0.3in}  N(\phi)= \int_{\phi_f}^{\phi_i}3\phi^2\frac{\left[\left(V_1\gamma^2({V_0\over V_1}-{1\over \phi})+\sqrt{V_1^2\gamma^4({V_0\over V_1}-{1\over \phi})^2+32\gamma^3 f_0^3}\right)^{2\over 3}-3.175\gamma f_0\right]}{1.26\gamma V_1\Big[V_1\gamma^2({V_0\over V_1}-{1\over \phi})+\sqrt{V_1^2\gamma^4({V_0\over V_1}-{1\over \phi})^2+32\gamma^3 f_0^3}\Big]^{1\over 3}}d\phi.\end{split}\end{small}\ee

Again despite such huge structure of equations, we are able to present a table of data set for the  expressions \eqref{epsaN}. In Table-3, we have varied $ V_1$ between $4.5~M_P^{5}\leq V_1 \leq   5.0 ~M_P^{5}$, so that $r$ and $n_s$ lie within the experimental limit. Further, the number of e-folds for Table-3 is found to vary with in the range $40 \leq N \leq 45$, which is more-or-less sufficient to solve the horizon and flatness problems. Clearly, the agreement with the observational data is outstanding, since the tensor to scalar ratio is able to sustain further constraints, which might appear from future analysis.\\
\begin{figure}
\begin{minipage}[h]{0.97\textwidth}
      \centering
      \begin{tabular}{|c|c|c|c|c|c|}
     \hline\hline
      $\phi_f$ in $M_P$ &$ V_1$ in $\times 10^{-6}~ M_P^{5}$ & $n_s$ & $r$ & $ {N}$ &$ {H_*}^2$ in $\times 10^{-6}~M_P^{2}$\\
      \hline
        0.75450 &4.5 &  0.96656 & 0.005836 & 40& 1.33\\
        0.74478 &4.6 &  0.96742 & 0.005593 & 41& 1.37\\
        0.73540 &4.7 &  0.96823 & 0.005364 & 42& 1.40\\
        0.72637 &4.8 &  0.96899 & 0.005149 & 43& 1.43\\
        0.71750 &4.9 &  0.96974 & 0.004947 & 44& 1.47\\
        0.70900 &5.0 &  0.97044 & 0.004756 & 45& 1.50\\
\hline\hline
    \end{tabular}
     \captionof{table}{Data set for the inflationary parameters in the case of $F(\mathrm T)=f_0 \mathrm T + f_1 (\mathrm T)^{3}$, taking into account $\phi_i=4.0~M_P$, ${V_0}=2.0 ~M_P^{4}$,~$\gamma=5\times  10^{7}~M_P^{-2}$, ~$f_0=0.5 ~M_P^2$ varying ${V_1}$ between $4.5\times 10^{-6}$ to $5.0\times 10^{-6}~ M_P^{5}~$.}
      \label{tab:Table3}
   \end{minipage}
   \end {figure}

The energy scale of inflation (${H_*}$) is shown in the last column of Table-3, which is of the order $10^{-3 }~M_P$ i.e. in the sub-Planckian scale, while the energy scale of inflation in a single scalar field model corresponding to GTR \cite{28} is given by the following expression,

\be\label{HE}{H_*} = 8\times 10^{13}\sqrt{\frac{r}{0.2}}GeV = 1.28\times 10^{13} GeV \approx 5.25\times 10^{-6}~M_P,\ee
whose numerical value is computed taking into account the value of the tensor-to-scalar ratio $r = 0.005149$ from the data set of Table-3.\\

Finally, to handle the issue of graceful exit from inflation, we recall equation \eqref{E2}, which in view of the above form of the potential, $V(\phi)=V_0-{V_1\over \phi}$, is expressed as,

\be \frac{\gamma H^6}{V_1} + \frac{6f_0 H^2}{V_1}+\left[{\dot\phi^2\over 2V_1} - \left({V_0\over V_1}-{1\over {\phi}}\right)\right]=0.\ee
During inflation, $H^2$ and $V_1$ are of the same order of magnitude, while the Hubble parameter varies slowly. But, at the end of inflation, the Hubble rate decreases sharply, and $\gamma H^6$ falls much below $V_1$. Hence, one can neglect both the terms ${\gamma H^6\over V_1}$ and ${f_0 H^2\over V_1}$  without any loss of generality. In the process one obtains,

\be{\dot{\phi}}^2=-2\left[V_0-{V_1\over \phi}\right].\label{osc2}\ee
Taking into account the values: $\phi_f=0.70900~M_P$, $V_0= 5.0\times 10^{-6}~M_P^4$ and $V_1= 2.0\times 10^{-6}~M_P^5$ from Table-3, it is possible to show that the above equation exhibits following oscillatory behavior,

\begin{align}\phi=\exp({i\omega t}),\end{align}
provided, $\omega \approx 1.476 \times10^{-3}~M_P$.
Hence, graceful exit is also exhibited. In a nutshell, both the forms $F(\mathrm{T})$ of the generalized metric teleparallel gravity are found quite susceptible to explain the cosmological evolutionary history, right from the very early inflationary era till date.

\subsection{Symmetric Teleparallel $F(Q)$ theory of gravity:}

Although everything appears to be nice, non-trivial $F(\mathrm T)$ theories are not locally Lorentz invariant, the most fundamental symmetry in physics, resulting in extra degrees of freedom not present in GTR. In fact, the lack of locally Lorentz invariance requires to deal with 16 equations, instead of the 10 associated with GTR. Einstein suggested to identify these extra degrees of freedom with the electromagnetic fields. However, he failed to find a consistent tensor-like description of the electromagnetic field equations in this approach. Further, while studying perturbations on the top of RW background, models based on $F(\mathrm T)$ gravity appear to suffer from strong coupling problems. In particular, if the action is expressed in the Einstein's frame, and perturbation is studied around the flat Minkwoski space-time, the spectrum of `general theory of relativity' reappears, being free from all the additional degrees of freedom. That is, the genuine physical degrees of freedom lose their kinetic term at the quadratic order, resulting in the break down of the standard perturbation theory. Since no new scalar mode is present in linear and second order perturbations in $F(\mathrm T)$ gravity, a strong coupling problem appears. Additionally, $F(\mathrm T)$ theory gravity can not make any prediction for `cosmic microwave background radiation' and `large scale structure' within the standard RW model due to strong coupling \cite{3}. Therefore $F(Q)$ gravity is currently in the lime-light, since these problems do not arise in $F(Q)$ theory of gravity, as already mentioned. This led people to consider generalized symmetric teleparallel $F(Q)$ gravity theory. Recently $F(Q)$ theory has been studied largely in different perspectives \cite{29,31,32,33,34,35,36,37,38}. Further, the field equations in view of the variational principle have also been derived \cite{8}. In the $F(Q)$ theory, one may use the special FLRW metric in Cartesian coordinates and the coincident gauge in this setting, which make the calculation easier by reducing the covariant derivative to merely a partial derivative. In the coincidence gauge, all the connections are made to vanish globally in a set of local Lorentz frame. As a result, the affine connection vanishes everywhere in a manifold and also both the Riemann-tensor ${\tilde R^\alpha}_{\beta\mu\nu}$ as well as the torsion-tensor $\mathrm T_{\alpha\mu\nu}$ vanish globally. In the coincidence gauge the non metricity scalar reads as $Q=-{6\dot a^2\over a^2N^2} = -6H^2$, under the gauge choice $N = 1$. Consequently, the Friedmann equations become identical to those of the $F(\mathrm T)$ gravity theory, and all the results found in connection with $F(\mathrm T)$ gravity theory holds, averting the problems associated with $F(\mathrm T)$ gravity theory.\\

Unfortunately, as soon as one considers a nonlinear generalization of STEGR, i.e., $F(Q)$ theory of gravity, it is also found to run from some serious pathologies. Covariant formulation of the theory requires to introduce St\"{u}ckelberg fields, in view of which the presence of ghosts induces Ostrogradski's instability. In the recent years it is shown that there is a strong coupling problem with the scalar perturbations around maximally symmetric (De-Sitter and Minkowski) backgrounds and also there might be a potentially strong coupling problem in the vector sector for flat cosmology \cite{4}. The strong coupling issue in the maximally symmetric Minkowski background mostly tell upon at very small scales, such as when applying it to the Black-holes in the asymptotically Minkowski region. Further, lack of diffeomorphic invariance in the self-interaction of graviton might possibly gives rise to Boulware-Deser ghost \cite{40}. Particularly, while studying cosmological perturbation around the spatially flat $F(Q)$ gravity theory, seven degrees of freedoms  are found propagating in the gravitational sector, out of which at least one is ghost \cite{41}. Nonetheless, The isotropic and homogeneous Robertson-Walker metric \eqref{RW} under present consideration also admits three non-trivial connections \cite{8,42}, apart from the coincidence gauge. Study of the other three connections have been initiated recently \cite{43,44}. More recently, it has been shown that all these three connections can also explain the cosmic evolutionary history, for a linear form of $F(Q) = f_0 Q$, without further modification \cite{45}. Since no generalization to non-linearity is required, these might be free from pathologies.\\

\subsection{Palatini formalism:}

For the sake of completeness, let us add a few lines in the context of yet another theory of gravity presented by Palatini long ago, which again admits late-stage of cosmic acceleration upon suitable modification \cite{44a}. In fact, while the metric $g_{\mu\nu}$ measures the distances in manifold and angles in tangent space, the connection ${\Gamma^\alpha}_{\mu\nu}$ actually maps between different tangent spaces that define parallel transport. Once the physical consideration that the principle of equivalence and causality should be related to each other is demanded, one enters into the realm of GTR, and ${\Gamma^\alpha}_{\mu\nu} = \{_\mu{^\alpha}_\nu\}$, the Levi-Civita. On the contrary, Palatini formalism treats metric and the connection as independent quantities. Hence, mathematically Palatini formalism appears to be more rigorous and enriches the geometric structure of the spacetime. The generalized action being function of both the metric and the connection is given by,

\be A(g, \Gamma) = \frac{1}{16\pi G}\int\sqrt{g} d^4 x F(\mathfrak{R}) + S_m,\ee
where $\mathfrak{R}$ stands for Ricci scalar in Palatini formalism and $S_m$ is the matter action. Note that $\delta {\Gamma^\alpha}_{\mu\nu;\alpha} - \delta {\Gamma^\alpha}_{\mu\alpha;\nu}$ and since there is no way to find the connections, the Ricci scalar $\mathfrak{R}$ remains unknown. The field equations are,

\be \label{PF1} F'(\mathfrak{R})\mathfrak{R}_{\mu\nu} - \frac{1}{2}F(\mathfrak{R})g_{\mu\nu} = \kappa T_{\mu\nu},\ee
whose trace yields
\be\label{PT}  F(\mathfrak{R}) \mathfrak{R} - 2 F(\mathfrak{R}) = \kappa T.\ee
One can immediately notice that for traceless fields $T = 0$, such as in the vacuum ($p = \rho = 0$) and in the radiation ($p = {1\over 3}\rho$) eras, in the context of cosmology, the above equation \eqref{PT} yields $F(\mathfrak{R}) \propto \mathfrak{R}^2$. As $\mathfrak{R}$ is obscure, so one does not have any idea regarding the evolution. Nonetheless, the evolution, whatever it might be, is identical in both the eras, which is physically unrealistic. On the other hand, for $F(\mathfrak{R})= \alpha \mathfrak{R} + \beta \mathfrak{R}^2$, equation \eqref{PT} for traceless fields results in $\mathfrak{R}^{n-1} = \frac{\alpha}{\beta(n-2)}$, for $n \ne 1,2$. Thus $\mathfrak{R}$ is a constant and de(anti)Sitter solution emerges (we shall make it clear in the following) in both the vacuum and radiation eras, which is unrealistic as already mentioned. Clearly one requires to consider additional fields, such as a scalar field to obtain a physically reasonable outcome. But, prior to that, it is somehow required to relate $\mathfrak{R}$ with $R$, so that one can study cosmological evolution. This is possible under the conformal transformation, $g_{\mu\nu} = F'(\mathfrak{R}) \tilde g_{\mu\nu}$, whence the action takes the following form \cite{44b},

\be \label{PC} A_P = {1\over 2\kappa} \int \sqrt{-\tilde g}d^4x[\Phi R - {3\over 2\Phi}\partial_\mu \Phi \partial^\mu\Phi - U(\Phi)] + S_m,\ee
where,
\be \label{R} \mathfrak{R} = R - {3\over 2\Phi^2} \partial_\mu \Phi \partial^\mu\Phi - {3\over \Phi} \Box\Phi,~\Phi = F'(\mathfrak{R}),~U(\Phi) = \Phi\chi(\Phi) - F(\chi)\ee
and the new field $\chi = \mathfrak{R}$. In the linear theory $F(\mathfrak{R}) \propto \mathfrak{R}$, $\Phi$ becomes a constant, so is $U(\Phi)$, as a result $\mathfrak{R} = R$, in view of \eqref{R} and GTR is retrieved. Phase-space structure of the scalar-tensor equivalence of Palatini action has been formulated, quantized and it was found that the semi-classical wavefunction is classically forbidden, which is a serious pathology of the formalism. Next, it was observed that either the scale factor evolves exponentially both in the early vacuum dominated and radiation dominated eras, which is unrealistic, or it admits Friedman-like solution $a \propto \sqrt t$ in the radiation era, but unfortunately a power law inflation is administered in the vacuum era without graceful exit. The fact that Palatini formalism runs into contradiction with the Standard model of cosmology was also unleashed earlier \cite{44c}. Further, in an attempt to analyze the stellar structure it was revealed that,  the formalism exhibits a singular behavior, giving rise to infinite tidal forces on the surface\cite{44d}. Thus Palatini formalism lacks novelty.

\section{Concluding remarks:}

In this article we have reviewed some earlier works with suitable modification focussing on the reconstruction program of the forms of modified and generalized teleparallel gravity theories. The importance of the work is, all the forms presented here admit a viable radiation and early matter dominated eras, which are at par with the standard (FLRW) model of cosmology. Further, these models can also explain the late-time cosmic accelerated expansion. Since inflation in the context of modified $F(R)$ theory of gravity has been studied widely, we have not considered it here. However, for the two possible forms of the teleparallel $F(\mathrm{T})$ gravity, we have studied inflation and found that the inflatioanry parameters are quite in agreement with the recently released observational data. Also since symmetric telepallel $F(Q)$ gravity theory in coincidence gauge, produces identical field equation as those for $F(\mathrm{T})$ gravity theory, so all the results obtained for $F(\mathrm{T})$ gravity holds for coincidence general relativity also.\\

While, the results obtained here are quite encouraging, nonetheless, all these theories suffer from some sort of unavoidable pathologies, as discussed in the literature. In the mean time, linear and second order scalar perturbations and the strong coupling issues of $F(\mathrm{T})$ gravity have been reinvestigated applying the effective field theory \cite{45}. Of course, no new scalar mode is found suggesting strong coupling issue. Nonetheless, authors have also shown that this can be avoided at some scale comparable with the cutoff scale of the applicability of the theory \cite{45}. Clearly nothing has been settled as yet. In this context, it may be mentioned that in the RW metric \eqref{RW} under consideration, symmetric teleparallel gravity theory admits four different connections. Apart from the coincidence gauge, the other three involve an additional non-vanishing function of time $\gamma(t)$, and as a result some difficulty arises in studying these connections. Nonetheless, very recently, we have prepared a detailed study with other three connections (communicated), which exhibit cosmic evolution at the linear level, without any modification. It should also be mentioned that, Bianchi identity, which is an outcome of the invariance of curvature by isometries of the metric tensor, holds automatically for GTR (${G^{\mu\nu}}_;\mu =0$) and is also true for $F(R)$ gravity, resulting in classical conservation of the energy momentum tensor. Nonetheless, this does not hold in general for teleparallel gravity theories. Even so, in the background of isotropic and homogeneous Robertson-Walker metric, Bianchi identity holds for $F(\mathrm T)$ gravity uniquely. In the case of $F(Q)$ theory of gravity, coincidence gauge ($Q = -6H^2$) is usually taken into consideration, in which case also Bianchi identity holds automatically. Nevertheless, for other connections it was not known with certainty. It has been exhibited \cite{46} that apart from one single situation, it holds in general for all the connections. We have also briefly discussed Palatini formalism, which appears to be a non-viable alternative to GTR.

\end{document}